# James Fergason, a Pioneer in Advancing of Liquid Crystal Technology

**Amelia Carolina Sparavigna**
Department of Applied Science and Technology, Politecnico di Torino, Torino, Italy

James Lee Fergason (1934 - 2008) focused his research on the liquid crystals. His studies correspond to a relevant part of the history of soft matter science and technology of liquid crystals. Here a discussion of some of his researches.

## 1. Introduction

The study of liquid crystals is pervasive of many areas of science and engineering. Besides fundamental studies on their physical and chemical properties, the researches on applications of this large family of materials are still continuing, providing new solutions to several technological problems. Their most common application is in the technology of liquid crystal displays (LCDs), fundamental for our social life: today, LCDs are so ordinary that nobody is taking care of the displays in mobile phones or of the flat desktop monitors and notebooks.

Among the pioneers of the research on liquid crystals and their applications, we find James Lee Fergason [1]. From November 1957, when he saw his first liquid crystal, he started his successful work on such materials and on the displays that we can have from them. James Fergason received more than 150 U.S. patents on liquid crystal technology, leading to commercial products such as medical and safety devices and displays for watches. On 2006, Fergason, was honoured with the Lemelson-MIT Prize, the largest cash prize given in the United States for invention [2].

Fergason (1934 - 2008) was born on a farm near the small town of Missouri. After the graduation from Carrollton High School in 1952, he enrolled at the University of Missouri, earning a Bachelor's Degree in physics in 1956. After serving in the Army, James Fergason was hired by Westinghouse Research Laboratories in Pennsylvania. There, he began groundbreaking work with cholesteric liquid crystals, forming the first industrial research group into the practical uses of the liquid crystal technology. After this work, Fergason earned his first patent (U.S. Patent 3114836A [3]), published in 1963, a patent on the use of thermochromic liquid crystals. Among the later uses of his research for the first patent we find liquid crystal thermometers and even a 1970s pop culture icon, the "mood ring".

Fergason's researches correspond to a relevant part of the history of the soft matter science and technology of liquid crystals. Let us follow his researches starting from patent on cholesterics.

## 2. Thermochromic liquid crystals

US Patent 3114836A is about an invention related to thermally responsive devices, particularly to those devices able to detect thermal patterns and convert them to visible patterns. It is then a patent about a peculiar display showing an image related to a temperature field. The core of the display is a thin film of cholesteric liquid crystals which exhibits "a property, upon interaction with light, which may be termed selective scattering. – tells the patent – A selectively scattering film, when observed with light impinging the film on the same side as that which is viewed, has an apparent color which is the complement of the color of the light transmitted by the film. Most materials do not selectively scatter light in that the light transmitted through them is not complementary to the light scattered by them but rather is within the same wavelength band. The phenomenon of selective scattering as exhibited by cholesteric liquid crystalline films is independent of whether the light illuminating the film is polarized or not. The color and intensity of the scattered light depends upon the temperature of the scattering material and upon the angle of incidence of illumination." Then, the patent continues with other properties of cholesterics films.

Let us remember that liquid crystals are anisotropic materials, which have some fluid phases that exist between the solid phase and the conventional isotropic liquid phase. Among the liquid crystals, we have the thermochromic liquid crystals (TLC), which are materials suitable for the temperature visualization. They are cholesterics, that is, chiral-nematic liquid crystal materials able to reflect definite colors at specific temperatures [4,5]. The color change for the TLC ranges from clear at ambient temperature, through red as temperature increases and then to yellow, green, blue and violet before turning colorless (isotropic) again at a higher temperature [5]. Therefore, the material appear colorless above and below the active range. The color-temperature play interval depends on the TLC composition. These color changes are repeatable and reversible as long as the TLC's are not physically or chemically damaged. They can be painted on a surface or suspended in a fluid and used to indicate visibly the temperature distribution. And therefore they can have several applications also in the field of microscopic thermal management [5].

The cholesteric nematics are twisted nematics: they are able of reflection of certain wavelengths by the microscopic order of the anisotropic chiral mesophase. Only this mesophase has thermochromic properties. The twisted nematic phase is a chiral mesophase where the molecules are oriented in layers with regularly changing orientation, which gives them a periodic arrangement. The light passing through the material undergoes Bragg diffraction on these layers and the wavelength with the greatest constructive interference is reflected producing a spectral color. When the temperature change, the spacing between the layers is changed and so the reflected wavelength. As previously told, the color of the thermochromic liquid crystal can continuously range through the spectral colors, depending on the temperature.

The TLCs used in dyes and inks are often microencapsulated, in the form of suspension. For instance, the H.W. Sands' pigments contain microencapsulated cholesteric liquid crystal mixtures that react to changes in temperature. Some TLCs are provided in a water-based ink formulation for application to polymeric substrates, such as thermometers for room, refrigerator, aquarium, and medical uses. A popular application was in the "mood rings": in 1975, jewelry designer Marvin Wernick had the idea of using TLCs for creating a ring. He used a hollow glass shell, filled with a thermotropic liquid crystal, instead of the stone of a ring: when worn, the TLC would change temperatures and color according to the temperature of the skin. Wernick failed patenting this ring. Two others inventors, Josh Reynolds and Maris Ambats, proposed bonded liquid crystals with quartz stones into rings; however, they too missed the opportunity "to ride the wave of the new fad". Mood rings became very popular during the 1970's and for a few years, were considered a piece of jewelry; they are now seen as an icon of the 1970s pop culture [6,7].

**3. Fergason at Westinghouse**
In 1965 and 1966, Fergason published some papers on cholesteric liquid crystals [8-11]. In the report of 1965 [11], the authors described the properties of temperature sensitive cholesteric liquid crystals proposing them for large area displays. In the Westinghouse laboratory, "an experimental 10 in. x 12 in. display panel was built to demonstrate the color, resolution and time constant capabilities of liquid crystals. Five different methods for obtaining dynamic thermal patterns in response to video signals are described. The application will dictate which method should be used for any given display problem. Considerable experimental work has been carried out on those methods which appear to be of most practical usefulness." [11] At the same time, Fergason and the Westinghouse Laboratory provided samples of liquid crystals mixtures to other researchers as reported in Refs.12 and 13. Fergason published also a paper [14] showing some preliminary studies by the authors, indicating that a liquid crystal film provided a demonstration and quantification of skin temperature patterns.

Liquid crystals were also used for the detection of gases and X-rays. such as in Ref.15. A set of reactive liquid crystal materials were developed which are capable of detecting quantities (10 ppm or less) of HCl, HF, hydrazine, UDMH, and nitrogen dioxide. The materials exhibit a change in color transition temperature upon exposure to these contaminant gases or vapours. Except for HF and HCl, the contaminants are readily distinguished from each other, as told in [15].

In US 3657538 A patent [16], it is proposed a device for detecting x-ray radiation using a cholesteric detector. "Increasing dosages of X-ray radiation progressively lower the color-play range of cholesteric-phase liquid-crystal materials. The effect is enhanced when an effective amount of an iodine-containing compound is used in the liquid-crystal material. Novel iodine-containing compounds are described, and articles are disclosed that give direct-reading indication of the dosages of X-ray radiation that they have received, without need for a separate development operation." The invention "is based upon observing the degree to which a shift in the color-play temperature range of a cholesteric-phase liquid-crystal material has taken place. … Cholesteric-phase liquid-crystal materials containing iodine display this shift more markedly than other known cholesteric-phase liquid-crystal materials do, so that the use of such compounds is preferred whenever greater sensitivity is desired."

The research for new devices using of liquid crystals in X-ray detectors is still alive, as shown by Ref.17. As told in this paper, the recent study of digital X-ray detectors in medical diagnostics has focused on high-resolution image acquisition. The reference is proposing a new structure for a non-pixel detector by constructing multiple layers, including photoconductor and liquid crystal (LC) cell layers. The results of this study indicate that an LC-based non-pixel detector is feasible for application in digital X-ray systems.

**4. The twisted nematic cells**

From 1966 to 1970 Fergason was Associate Director of the Liquid Crystal Institute at Kent State University. As told in Ref.1, during this period he was "part of an effort to use liquid crystals for thermal mapping to screen for breast cancer". At the same time, Fergason became "an early pioneer in developing electronic display applications and increasingly focused his attention on nematic materials." And in fact, in 1969, his researches produced a strongly improvement of controlled nematic cells. Let us remark that already in 1965, he filed for a patent on electric field devices based on cholesterics liquid crystals [18].

According to the IEEE [19], between 1964 and 1968, at the RCA David Sarnoff Research Center in Princeton, a team of engineers and scientists led by G. Heilmeier, L. Zanoni and L.Barton, devised a method for an electronic control of light reflected from liquid crystals. The liquid crystals displays were based on the dynamic scattering method (DSM): an electrical charge was applied to rearrange the molecules to scatter light. However, the DSM cell had a poor contrast and required a large amount of energy, therefore the researchers on liquid crystals were searching for other effects, to replace DSM. And then this cell was soon replaced by cells based on the use of the twisted nematic field effect of liquid crystals, the TN-LCDs.

Reference 20 is telling that "one of the most important major breakthroughs occurred in the late 1969 when J.L. Fergason, working at the newly formed firm, International Liquid Crystal Company (ILIXCO) in Kent, Ohio, discovered the twisted-nematic field-effect LCD." This effect was successful to create displays for watches, calculators, and later, other applications including large screens. "Because Mr. Fergason's patent application was not made public until several years later, Drs. Wolfgang Helfrich and Martin Schadt of F. Hoffman LaRoche in Basel, Switzerland, published a paper on the same effect in 1971 and were awarded a patent in 1975. Needless to say, this sparked a long legal battle over ownership of the invention. Eventually the issue was settled out of court. That Mr. Fergason is generally regarded as the inventor of the TN-LCD is exemplified by the fact that he was awarded the highest honor of the Society of Information Display for the initial discovery." Fergason obtained US3731986 patent for "Display Devices Utilizing Liquid Crystal

Light Modulation" [21]. In 1971, the International Liquid Crystal Company (ILIXCO) owned by James Fergason produced the first modern LCD watch based on it.

As told in [22], it was in August 1968, that Fergason described the concept of helical twist and rotation of linearly polarized light [23]. In January 1970, Fergason reported the concept of reorienting nematic liquid crystals by applying an electric field [24]. In the Reference 22, the author is describing his interaction with James Fergason: "When I asked Fergason in 1995 how he could have applied for a patent in April 1971 if he had already published the idea in 1968 [22] and in 1970 [24], he showed me a nonpublic document 'Liquid crystal nonlinear light modulators using electric and magnetic fields.' The document contained a full account of the concepts of the TN mode. The document said: 'The invention was conceived on December 30, 1969. The invention was explained to Ted R. Taylor and Thomas B. Harsch the same day. The invention was first reduced to practice on April 5, 1970.' Taylor and Harsch, who countersigned the document, were the coauthors of his 1970 paper [24]. He said that the examiner at the U.S. Patent Office recognized the importance of the descriptions in the electro-technology paper. The patent application of April 1971 was granted in the U.S. and also in Germany." [22]

In 1970, Fergason published several papers. One of them is, in my opinion, important in this discussion on twisted nematics. In the chapter entitled, Liquid-Crystal Detectors, in a book on Acoustical Holography [25], Fergason is discussing the twisted cell. He explains that, if we define a plane of area A of nematic uniformly aligned, which is separated from a plane of material oriented at an angle $\varphi$ with respect to the first plane and separated by a distance Z, then a torque will be exerted which is given by $T=(A\varphi/Z)k$, where k is a force constant whose units are dynes. Associated with this torque we have an energy $E=(A\varphi^2/2Z)k$. The constant k was estimated to be $2 \times 10^{-7}$ dyn. As Fergason remarks, "this gives us insight into the nature of detectors which are made with liquid crystals. Because of the extremely small force constant, the orientation can be effected by a very small force." He estimates that the free energy in a sheet 0.1 mm thick, with a twist of 90° is $2 \times 10^{-13}$ W sec/cm$^2$, and this is far above a threshold effect. Therefore it is reasonable to modulate a liquid crystal system with an input power of $10^{-14}$ W/cm$^2$. Fergason continues telling that the reorientation which takes place in nematic phases is detected by the use of cross-polarizers. The paper [25] is a clear description of the twisted nematic effect. Fergason tells also that "the nematic phase which is twisted gives us insight into the cholesteric phase. If we were able to twist a nematic phase beyond 90° and continue to twist until the material twisted 90° in one-half the wavelength of visible radiation, we could have the cholesteric phase." This is the idea of a super-twisted nematic cell.

Let us also remember the paper by J.L. Ericksen [26], where he is telling that "In a lecture given at Johns Hopkins, Dr. J. L. Fergason discussed preliminary observations concerning orientation waves which propagate, inducing little or no motion of the fluid. Theoretically, the observed waves should induce some motion, but we here discuss an exceptional type of wave which need not." For what concerns the Kawamoto's question about the patent, let us consider that Fergason experimented in general on the twisted nematic effect, publishing some papers, but that he patented an electronic device to drive the effect, able to switch the cell from the twisted state to the uniform state driven by an electric field.

The optical behavior of twisted nematics had first been noticed by French physicist Charles-Victor Mauguin in 1911 [27]: he observed that when a liquid crystal is sandwiched between two aligned polarizers, he could twist them in relation to each other, but the light continued to be transmitted. Normally if two polarizers are aligned at right angles, light will not flow through them. Mauguin concluded that the light was being repolarized by the twisting of the crystal itself [27]. Applying an electric field, the molecules of the liquid crystals are oriented by the field and then they assume a uniform alignment, perpendicular to the cell walls. Therefore, between crossed polarizers, the materials is assuming the behavior of a uniform liquid and therefore we have the light extinction. The Figure 1 is schematically showing a TN cell.

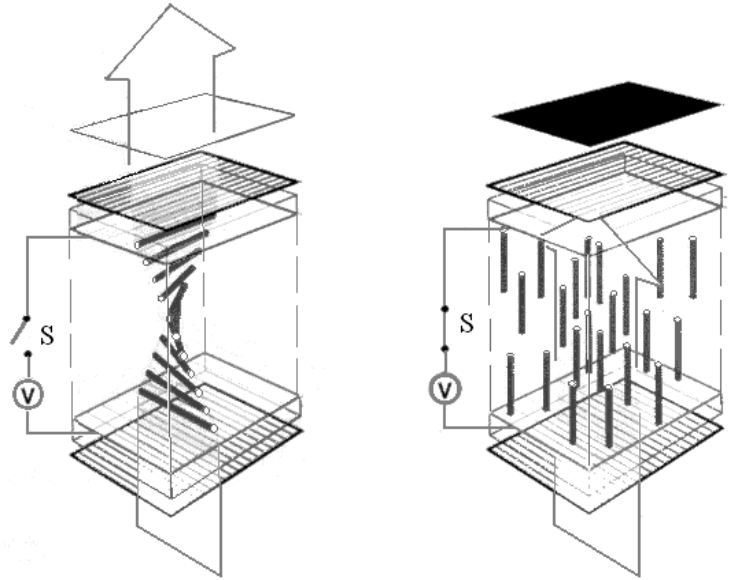

**Figure 1** - Illustration of operation of a pixel in a twisted nematic cell. On the left, no voltage is applied. The nematic is twisted and the light is passing in the cell with crossed polarizers. When a voltage V is applied (right diagram), the nematic assumes a uniform alignment which is seen by the light as an isotropic medium, therefore the light is stopped by the crossed polarizers.

Twisted nematic liquid crystal displays were superior to the earlier dynamic scattering displays, and soon became widespread. ILIXCO Fergason's company began manufacturing and marketing the new displays in 1971. His first customers were the Bulova Watch Company and Gruen Watch Company which used the technology to market the first LCD watches using this technology. At the same time, Wolfgang Helfrich and Martin Schadt were also working on twisted nematics and published a paper on the same effect in 1971 [28]. The Ref.22 describes in detail the works on twisted nematics made by Helfrich.

**5. Smectic liquid crystals**
The smectic liquid crystals were first identified merely as 'slimy liquid crystals' by Otto Lehmann in the late 1880s, and then identified as layered phases by Georges Friedel in 1922 [29,30]. There are two principal phases, the smectic A phase, in which the liquid crystal molecules are oriented normal to layers, and the smectic C phase in which they are tilted.
One of the first papers on the smectic C phase is that written by H. Arnold [31], on the heat capacity and enthalpy of phase transitions. Arnold used adiabatic calorimetry. He made for the first time an especial reference "to the transition between smectic modifications of different types." He compared the enthalpy for the smectic B to smectic A transition and for the smectic C to smectic A transition of two liquid crystals. In a Google Scholar search on the smectic C phase, after this paper of 1966 we find those of James Fergason. In fact, when he was at the Liquid Crystals Institute of Kent, he participated the initial studies of the smectic C liquid crystal phase [32-35]. For instance, in [32], we have that the researchers studied the new phenomenon of a temperature dependent tilt angle in the smectic C phase. The measurement of the tilt angle had been made using conoscopic observation and circularly polarized light. The material under investigation was the terephthal-bis-(4-n-butylanaline) that has nematic, smectic A, smectic C, and smectic B liquid-crystalline phases. The tilt angle of smectic C changes from 0° at the smectic-A-smectic-C transition to approximately 26° at the smectic-C-smectic-B transition temperature.
In [33], the authors describes convergent light observations on three liquid crystalline substances with nematic and smectic phases. "The nematic phase was observed to be uniaxial as expected, but smectic C was found to be biaxial." The biaxial character of the smectic C was partially explained by the anisotropy of the degrees of order.

**6. Encapsulated and Polymer Dispersed Liquid Crystals**
Let us continue using Google Scholar. If we search "JL Fergason" we find that his most cited publication is a patent of 1984 [36]. About Patent 4435047, entitled Encapsulated Liquid Crystal and Method: "according to one aspect of the invention, liquid crystal material, and especially nematic material, is encapsulated; according to another aspect the encapsulated liquid crystal material is used in liquid crystal devices, such as relatively large size visual display devices; and according to further aspects there are provided methods for encapsulating liquid crystal material and for making a liquid crystal device using such encapsulated liquid crystal material."
We have already told of the encapsulation of cholesteric for TLCs. Here instead of cholesterics, nematics are used. In the microencapsulation, the liquid crystal is mixed with a polymer dissolved in water: "A method for making encapsulated liquid crystals may include mixing together the encapsulating medium, the liquid crystal material (including, if used, the pleochroic dye material), and perhaps a carrier medium, such as water. Mixing may occur in a variety of mixer devices, such as a blender, a colloid mill, which is most preferred, or the like. What occurs during such mixing is the formation of an emulsion of the ingredients, which subsequently can be dried eliminating the carrier medium, such as water, and satisfactorily curing the encapsulating medium, such as the PVA." When the water is evaporated, the liquid crystal is surrounded by the polymer. A large number of tiny "capsules" are produced and distributed through the bulk polymer. The droplets produced in such a manner tend to be non-uniform in size. Materials manufactured by encapsulation are referred to as NCAP or nematic curvilinear aligned phase [37].
The NCAPs are then polymer-dispersed liquid crystals (PDLCs). Another way to prepare a PDLC is the phase separation [38]: the mixture is prepared by homogeneous solution of pre-polymer LC and curing agent / photo-initiator which catalyzes the polymerization process when the mixture is UV irradiated. Phase separation can also be thermally induced or solvent induced. Each method produces PDLCs with different properties and characteristics. Among the factors influencing the properties of the PDLC are the size and the shape of the droplets, and of course the types of polymer and liquid crystal used.
A PDLC display consists of droplets of liquid crystals inside a polymer network, as shown in Figure 2. These droplets are of micrometric size; inside them the molecules align themselves in a bipolar configuration, when the surface between the liquid crystal and the polymer is favoring a planar configuration. If the electrical field is zero, bipolar droplet axes have a random configuration, the axis direction is fixed by the droplet surface. When the electric field is above a threshold value, droplets are all oriented parallel to the field (Figure 2).

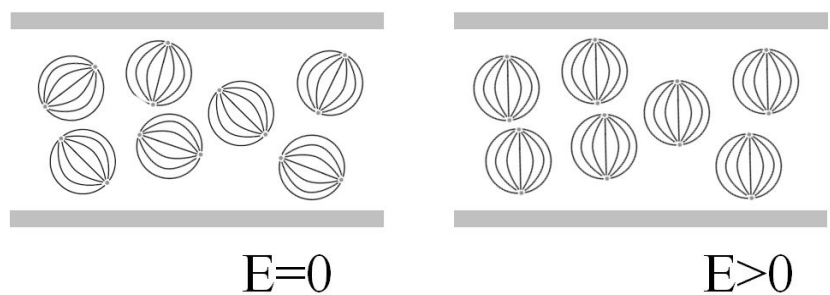

**Figure 2 - Bipolar droplets without and with an electrical field applied to the film: if *E=0*, bipolar droplet axes assume a random configuration but when the electric field is above a threshold value, droplets are all oriented parallel to the field.**

In the off state, the droplets are randomly aligned, the refractive index seen by the light is different from that of the polymer and then the light is scattered or reflected towards the viewer. In the on state, the LC molecules orient along the direction of the applied field: in this case the refractive index seen by the light is close to that of the polymer and then, due to this matching of indices, the light can be transmitted. The device does not contain polarizers.

PDLCs can be used in light valves for instance, however one of the common use is made into the privacy windows or for smart projection screens. The PDLCs are then the materials suitable for the smart glasses, or switchable glasses, where the glasses change light transmission properties when voltage is applied.

**7. Technology incubation**

Besides the several patents on the liquid crystals encapsulation, we have patents for other applications. Let me cite a patent [39], about an optical read/write information storage system, including a liquid crystal information storage medium. In this device, "the volumes of liquid crystal material being operative to modulate or not to modulate light, e.g. by transmitting or scattering light as a representation of logic 1 and logic 0 information, the storage medium being responsive to a first energy input to assume and to store one detectable output condition and to a second energy input to assume and to store a second detectable output condition, input energy systems for writing and erasing information, and an optical reading device for detecting such output conditions of the storage means as a representation of information stored therein." [39] There are also patents on liquid crystal projectors [40] and on devices for stereoscopic image selection [41].

As reported in [1], during the 1980s and 1990s, Fergason led a succession of self-funded research and technology incubation programs. Among his projects we have that of the already discussed plastic LCDs, that is, of the PDLC displays. James Fergason and his sons pioneered the market for eye protection devices too, based on LC shutters. "During the 1990s, Fergason filed a family of patents for technology and devices to overcome the commercial hurdles of personal viewers, sometimes called head-mounted displays, or near-to-eye displays. The inventions disclosed included optical dithering devices to increase imager resolution, the Retro-Vue™ head-mounted projector currently be evaluated as a military training device, and a range of viewer optics to enable multiple optical channels." [1]

In 2001, James Fergason founded the Fergason Patent Properties (FPP), with the assignment to the company of more than 35 issued patents. FPP is involved in several innovative products. Among the technologies there are the Dynamic Contrast Ratio and High Dynamic Range LCDs and the Stereo 3D Monitors. "Introduced in 2002, the System Synchronized Brightness Control (SSBC™) technology has been widely adopted by the leading LCD module and TV companies. The patented LCD video control system optimizes the backlight brightness and video signal to display a high contrast image with the intended brightness. The current list of licensees includes Sony, Sharp, Samsung, LG Display, JVC, Seiko, and others." [1]

**8. Conclusions**

We have seen that James Fergason strongly contributed to the advance of soft matter science, for his discoveries on the role of thermal and electric fields on the features of cholesteric and nematic liquid crystals. We have also discussed the several fundamental applications he proposed. Let us conclude this paper with an observation on the economic and social contributions of Fergason and, in general, of other persons who worked on liquid crystals. Again, Ref.1 is stressing the huge amount of money connected with the LCD component business. However, as it is told there for the United States, but here we can generalize to all the countries, by far the biggest contribution of the LCD industry to the global economy comes from the enabling of new information services. Thanks to the low power consumption and low price liquid crystals displays, the mobile handsets with integrated LCDs are the people portal to anywhere-anytime information [1].